\documentclass[12pt,letterpaper]{article}%
\usepackage{amssymb}
\usepackage{amsfonts}
\usepackage{amsmath}
\usepackage{hyperref}%
\usepackage{graphicx}
\usepackage{float}
\newtheorem{theorem}{Theorem}

\newtheorem{proposition}[theorem]{Proposition}

\DeclareMathSymbol{\mrq}{\mathord}{operators}{`'}

\begin{document}

\title{Newtonian Gravity on an N-Sphere}
\author{T. Curtright$^{1}$\thanks{{\footnotesize curtright@miami.edu}}\ \ and H.
Alshal$^{2}$\thanks{{\footnotesize halshal@scu.edu}}\medskip\\$^{1}\ $Department of Physics, University of Miami, Coral Gables, FL 33124\\$^{2}\ $Department of Physics, Santa Clara University, Santa Clara, CA 95053}
\date{}
\maketitle

\begin{abstract}
\textit{We consider some elementary features of Newtonian gravity, or
electrostatics, as defined on an N-sphere. \ In particular, we present and
discuss \textquotedblleft the shell theorem\textquotedblright\ for this
system.}

\end{abstract}

\subsection*{Introduction}

The analogue of Newtonian gravity on a closed Riemannian manifold has been an
interesting subject for quite a long time, especially among mathematicians
(e.g. see \cite{Aubin}\ and the literature cited therein). \ The N-sphere
presents a tractable example that can be analyzed without approximations (e.g.
see \cite{Chapling}\ and the literature cited therein). \ 

The physics of Newtonian gravity on the N-sphere is emphasized here.
\ Alternatively, one may think of the following as electrostatics on the
N-sphere. \ Either way, perhaps some insight and intuition can be gleaned from
this simple model. \ 

Taking this point of view, we follow a path not very far from that trodden by
G. Green so long ago \cite{Green}. \ We identify the point-particle
gravitational or electrostatic potentials with a Green function of the system,
and we define the force on such particles in terms of the gradient of that
potential, as is usual in physics. \ We then consider the analogue of Newton's
shell theorem for this simple system. \ We find a generalization of the shell
theorem which is simple to state as well as pleasing to our taste.

\subsection*{N-Sphere Laplacian Green Function}

On an N-sphere of radius $R$, acting on functions $f\left(  \theta\right)  $
with only polar angle dependence:
\begin{equation}
\nabla^{2}=\frac{1}{R^{2}}\left(  \frac{1}{\left(  \sin\theta\right)  ^{N-1}%
}\frac{\partial}{\partial\theta}\left(  \left(  \sin\theta\right)  ^{N-1}%
\frac{\partial}{\partial\theta}\right)  \right)
\end{equation}
For small angles, $\nabla^{2}\sim\frac{1}{R^{2}}\left(  \frac{1}{\theta^{N-1}%
}\frac{\partial}{\partial\theta}\left(  \theta^{N-1}\frac{\partial}%
{\partial\theta}\right)  \right)  =\frac{1}{R^{2}}\left(  \frac{\partial^{2}%
}{\partial\theta^{2}}+\frac{N-1}{\theta}\frac{\partial}{\partial\theta
}\right)  $. \ Also recall the \textquotedblleft surface
area\textquotedblright\ of the unit $N$-sphere $S_{N}$ is $A_{N}=2\pi^{\left(
N+1\right)  /2}/\Gamma\left(  \frac{N+1}{2}\right)  $. \ This is the
\textquotedblleft total solid angle\textquotedblright\ for Euclidean space
$\mathbb{E}_{N+1}$. \ 

A singular solution of the homogeneous equation
\begin{equation}
0=\frac{1}{\sin^{N-1}\theta}\frac{\partial}{\partial\theta}\left(  \left(
\sin^{N-1}\theta\right)  \frac{\partial}{\partial\theta}I\left(
\theta,N\right)  \right)
\end{equation}
is obviously given by the integral
\begin{equation}
I\left(  \theta,N\right)  =\int_{\pi/2}^{\theta}\frac{1}{\left(  \sin
\phi\right)  ^{N-1}}d\phi
\end{equation}
\emph{almost everywhere}, except at the singularities $\theta=0$ or
$\theta=\pi$. \ All $I\left(  N\right)  $ for even $N$ involve logarithms, but
all $I\left(  N\right)  $ for odd $N$ do not. \ 

Similarly, a singular solution of the inhomogeneous equation
\begin{equation}
1=\frac{1}{\sin^{N-1}\theta}\frac{\partial}{\partial\theta}\left(  \left(
\sin^{N-1}\theta\right)  \frac{\partial}{\partial\theta}J\left(
\theta,N\right)  \right)
\end{equation}
is just as obviously given almost everywhere by the double integral%
\begin{equation}
J\left(  \theta,N\right)  =\int_{\pi/2}^{\theta}\frac{1}{\left(  \sin
\vartheta\right)  ^{N-1}}\left(  \int_{\pi/2}^{\vartheta}\left(  \sin
\phi\right)  ^{N-1}d\phi\right)  d\vartheta
\end{equation}
By combining these two solutions we arrive at explicit expressions for a
Laplacian Green function on the $N$-sphere, in a form that exhibits the
physics of the model in simple terms. \ The result is%
\begin{gather}
G\left(  \widehat{r},\widehat{s},N\right)  =\frac{1}{2A_{N-1}}~I\left(
\arccos\left(  \widehat{r}\cdot\widehat{s}\right)  ,N\right)  -\frac{1}{A_{N}%
}~J\left(  \arccos\left(  \widehat{r}\cdot\widehat{s}\right)  ,N\right)
\label{GreenN}\\
R^{2}\nabla^{2}G\left(  \widehat{r},\widehat{s},N\right)  =\delta^{N}\left(
\widehat{r}-\widehat{s}\right)  -\frac{1}{A_{N}}\label{DiffEqn}%
\end{gather}
As usual in physics, $G\left(  \widehat{r},\widehat{s},N\right)  $ is
interpreted here as the potential at position $\widehat{r}$\ produced by a
unit point charge at position $\widehat{s}$.

It is straightforward to express the results (\ref{GreenN}) in terms of
hypergeometric functions \cite{Chapling}, which \emph{Mathematica} will do
without much coaxing, but the hypergeometric functions in question always
reduce to combinations of elementary functions. \ (See the Appendix for
$N\leq10$.) \ Alternatively, the Green function can be written as a sum of
bilinears in a complete set of hyperspherical harmonics \cite{Alshal} divided
by their Laplacian eigenvalues, but excluding the zero mode solution
\cite{Aubin}, hence the $-1/A_{N}$ term in (\ref{DiffEqn})

The Dirac delta produced by acting with $\nabla^{2}$\ at the \textquotedblleft
north pole\textquotedblright\ of the N-sphere is most easily exhibited by
small angle expansions. \ For example:%
\begin{align}
G\left(  \widehat{r},\widehat{s}=\widehat{z},N=2\right)   & =\frac{1}{4\pi}%
\ln\left(  1-\cos\theta\right)  =\frac{\ln\theta}{2\pi}-\frac{1}{4\pi}%
\ln2-\frac{\theta^{2}}{48\pi}+O\left(  \theta^{4}\right) \\
G\left(  \widehat{r},\widehat{s}=\widehat{z},N=3\right)   & =\frac{1}{4\pi
^{2}}\left(  \theta-\pi\right)  \cot\theta=\allowbreak-\frac{1}{4\pi\theta
}+\frac{1}{4\pi^{2}}+\frac{\theta}{12\pi}+O\left(  \theta^{2}\right)
\end{align}
etc. \ For $N>2$ the leading term is always $-1/\left(  \left(  N-2\right)
A_{N}~\theta^{N-2}\right)  $, thereby revealing the Dirac delta when the
Laplacian acts as $\frac{1}{R^{2}}\left(  \frac{\partial^{2}}{\partial
\theta^{2}}+\frac{N-1}{\theta}\frac{\partial}{\partial\theta}\right)  $ for
small $\theta$.

With $G\left(  \widehat{r},\widehat{s},N\right)  $ interpreted as a potential,
then the gradient of $G$ is to be interpreted as a repulsive/attractive force
exerted on a unit point charge/mass at position $\widehat{r}$ produced by an
identical point charge/mass at position $\widehat{s}$. \ Placing a source
point charge at the north pole\ of the N-sphere will produce a repulsive force
on another, same sign point charge at position $\left(  \theta,\cdots\right)
$. \ In this particular situation, the repulsive force depends only on the
polar angle $\theta$, is independent of all the other \textquotedblleft
azimuthal\textquotedblright\ angles, and has only a $\theta$ component as
given by
\begin{equation}
F\left(  \theta,N\right)  =\frac{d}{d\theta}G\left(  \theta,N\right)
=\frac{1}{4\pi^{N/2}}\left(  \Gamma\left(  \frac{N}{2}\right)  -\frac{2}%
{\sqrt{\pi}}~\Gamma\left(  \frac{N+1}{2}\right)  K\left(  \theta,N\right)
\right)  \frac{1}{\left(  \sin\theta\right)  ^{N-1}}%
\end{equation}
where $K\left(  \theta,N\right)  =\int_{\pi/2}^{\theta}\left(  \sin
\phi\right)  ^{N-1}d\phi=\left(  \sin\theta\right)  ^{N-1}\frac{d}{d\theta
}J\left(  \theta,N\right)  $. \ For example:

\begin{figure}[H]
  \centering
    \includegraphics[width=0.62\textwidth]{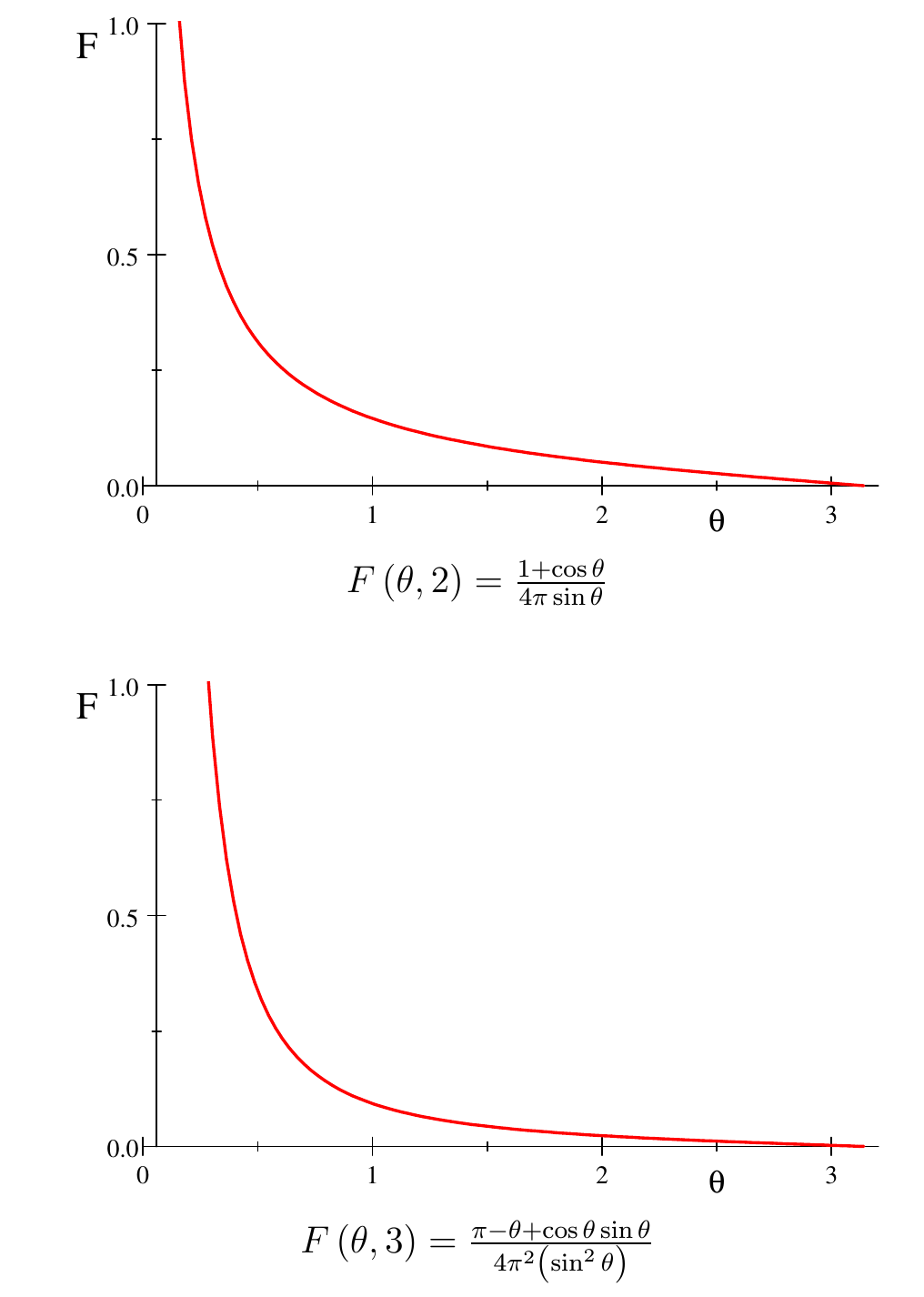}
\end{figure}

The result for $F\left(  \theta,N\right)  $\ evinces Gauss' law on the
N-sphere, as Faraday would surely have realized. \ The field line flux through
any $S_{N-1}$ subsphere at polar angle $\theta$, of radius $R\sin\theta$, is
$\Phi=\int F\left(  \theta,N\right)  R^{N-1}\sin^{N-1}\theta d^{N-1}%
\Omega=F\left(  \theta,N\right)  R^{N-1}A_{N-1}\sin^{N-1}\theta$. \ This flux
is proportional to the total charge on the hyperspherical cap $\left.
S_{N}\right\vert _{\leq\theta}$. \ Thus $\Phi$ decreases monotonically as
$\theta$ increases for $0<\theta<\pi$, and goes to zero as $\theta
\rightarrow\pi$ corresponding to zero total charge on the entire hypersphere.
\ For example:%
\begin{align}
\int F\left(  \theta,2\right)  \sin\theta d\phi & =\frac{1}{2}\left(
1+\cos\theta\right)  =1-\int_{0}^{\theta}\frac{1}{4\pi}\sin\vartheta
d\vartheta\int_{0}^{2\pi}d\phi\\
\int F\left(  \theta,3\right)  \sin^{2}\theta d^{2}\Omega & =\frac{1}{\pi
}\left(  \pi-\theta+\cos\theta\sin\theta\right)  =1-\int_{0}^{\theta}\frac
{1}{2\pi^{2}}\sin^{2}\vartheta d\vartheta\int d^{2}\Omega
\end{align}
etc. \ The \textquotedblleft$1$\textquotedblright\ on the RHS is the Dirac
delta contribution, while the integral is due to the uniform negative charge
density on the hypersphere.

\subsection*{The Shell Theorem}

An interesting physical feature of this N-sphere model is embodied in the
\textquotedblleft shell theorem\textquotedblright\ --- or perhaps the
\textquotedblleft new shell theorem\textquotedblright\ would be a more
appropriate description.

In Euclidean space the well-known shell theorem is the statement that an ideal
uniform spherical shell of charge or mass will produce no force on a test
particle placed within the shell. \ Actually, the theorem is true for general
closed, charged, equipotential shells (somewhat obviously from the point of
view of the uniqueness theorem) as was first proven for spheroidal shells by
Isaac Newton, in the \textit{Principia}, and much later for general
ellipsoidal shells by James Ivory, remarkably before potential theory was invented.

However, a uniformly charged $S_{N-1}$ sub-sphere located on the N-sphere
\emph{will} exert force on a test particle located at almost all points on
$S_{N}$.\footnote{More generally, on a generically curved, closed Riemannian
manifold, the interior of a charged, closed, equipotential submanifold is
\emph{not} necessarily at a \emph{constant} potential, despite what one might
naively expect from harmonic function theory. \ Recall the potential,
identified as a Green function, is \emph{not} necessarily harmonic in the
interior of the submanifold due to terms analogous to the $-1/A_{N}$ in
(\ref{DiffEqn}).} \ The only points on $S_{N}$ where the force on the test
particle will be zero are the \textquotedblleft two antipodal Faraday
points\textquotedblright\ of the system, as well as all points on the
uniformly charged $S_{N-1}$ itself. \ Note that the antipodal Faraday points
(or \textquotedblleft Far points\textquotedblright\ for brevity) may be
defined as the two points on $S_{N}$ for which all points on the embedded
$S_{N-1}$ are equidistant, with each \textbf{Far} point as \textbf{far} as
possible from the sub-sphere as well as from each other. \ These two Faraday
points maximize, locally, their respective distances from the $S_{N-1}$
sub-sphere. \ (Of course, to specify these Far points for the N-sphere,
distances may be computed either intrinsically or else extrinsically, if in
the latter case both the $S_{N-1}$ and the $S_{N}$ manifolds are canonically
embedded in $\mathbb{E}_{N+1}$.)

For such a uniformly charged or massive $S_{N-1}\subset S_{N}$, the shell
theorem is supplanted by the following (\textit{new shell theorem}): \ 

\begin{proposition}
The force on any test particle placed on $S_{N}\smallsetminus S_{N-1}$ is the
same as would result if all the uniformly distributed charge or mass on
$S_{N-1}$ were moved \textbf{away from} the test particle and relocated
entirely at the \textbf{opposing Faraday point}.
\end{proposition}

\noindent To clarify the last \emph{point}, note that the sub-sphere
partitions $S_{N}\smallsetminus S_{N-1}$ into two disjoint regions, with one
region containing the test particle and the other containing the
\textquotedblleft opposing\textquotedblright\ Faraday point. \ Thus the
position of the test particle determines which of the Far points is the
opposing one.

So stated, the spherical shell theorem in Euclidean space becomes a special
case of the Proposition. \ If a test particle is placed outside a uniformly
charged $S_{N-1}$ spherical shell embedded in $\mathbb{E}_{N}$, the force on
that test particle is the same as though all the charge were concentrated at
the center of the shell, as is well-known. \ On the other hand, if the test
particle is placed inside the shell, there is no force, but that is exactly
the same null result that would be obtained if all the charge on the shell
were moved out to the \textquotedblleft point at infinity\textquotedblright.
\ In this case, the center of the charged shell and the point at infinity play
the roles of the antipodal Faraday points\ for the system. \ That is to say,
the Euclidean space shell theorem for $\mathbb{E}_{N}$ results from taking the
radius of $S_{N}$ to infinity while keeping fixed the radius of the charged
$S_{N-1}$ sub-sphere. \ 

All this is most easily visualized and verified mathematically for the
2-sphere. \ 

\subsection*{The Simplicity of $N=2$}

For a 2-sphere of radius $R$, let $\overrightarrow{r}$ and $\overrightarrow{s}%
$ be two points on the sphere, so $r=R=s$. \ Then the standard Green function
for the Helmholtz equation is%
\begin{gather}
G_{\text{Helmholtz}}\left(  \overrightarrow{r},\overrightarrow{s}\right)
=\sum_{l=0}^{\infty}\sum_{m=-l}^{l}\frac{Y_{lm}\left(  \widehat{r}\right)
Y_{lm}^{\ast}\left(  \widehat{s}\right)  }{k^{2}-l\left(  l+1\right)  /R^{2}%
}=\sum_{l=0}^{\infty}\frac{2l+1}{4\pi}\frac{P_{l}\left(  \widehat{r}%
\cdot\widehat{s}\right)  }{k^{2}-l\left(  l+1\right)  /R^{2}}\\
\left(  \nabla^{2}+k^{2}\right)  G_{\text{Helmholtz}}\left(
\overrightarrow{r},\overrightarrow{s}\right)  =\sum_{l=0}^{\infty}\sum
_{m=-l}^{l}Y_{lm}\left(  \widehat{r}\right)  Y_{lm}^{\ast}\left(
\widehat{s}\right)  =\delta^{2}\left(  \widehat{r}-\widehat{s}\right)
\end{gather}
where $\widehat{r}$ has direction $\left(  \theta,\phi\right)  $,
$\widehat{s}$ has direction $\left(  \vartheta,\varphi\right)  $, and
$\widehat{r}\cdot\widehat{s}=\cos\Theta$, with
\begin{equation}
\cos\Theta=\cos\theta\cos\vartheta+\sin\theta\cos\phi\sin\vartheta\cos
\varphi+\sin\theta\sin\phi\sin\vartheta\sin\varphi
\end{equation}
and $\delta^{2}\left(  \widehat{r}-\widehat{s}\right)  =\frac{1}{\left\vert
\sin\Theta\right\vert }~\delta\left(  \Theta\right)  \delta\left(
\Phi\right)  $. \ Note that $\sqrt{1-\cos\Theta}$ is proportional to the
length of the chord (\emph{not} the arc length) that connects the two points
on the sphere. \ Unfortunately, \emph{Mathematica} cannot readily carry out
the sum over $l$ to obtain a simple closed form for $G_{\text{Helmholtz}}$,
even after regularizing the sum through use of an $i\epsilon$ prescription. \ 

This is in contrast to the Laplacian Green function on the 2-sphere. \ In this
case \emph{Mathematica} can perform the relevant sum.%
\begin{gather}
G_{\text{Laplace}}\left(  \overrightarrow{r},\overrightarrow{s}\right)
=\text{constant}+\sum_{\text{NB }l=1}^{\infty}\sum_{m=-l}^{l}\frac
{Y_{lm}\left(  \widehat{r}\right)  Y_{lm}^{\ast}\left(  \widehat{s}\right)
}{-l\left(  l+1\right)  /R^{2}}\nonumber\\
=\text{constant}+\sum_{l=1}^{\infty}\frac{2l+1}{4\pi}\frac{P_{l}\left(
\cos\Theta\right)  }{-l\left(  l+1\right)  /R^{2}}\nonumber\\
=\text{constant}+\frac{R^{2}}{4\pi}\left(  \ln\left(  1-\cos\Theta\right)
+1-\ln2\right)
\end{gather}
Note the issue with the $l=0$ mode\ (for example, see
\href{https://en.wikipedia.org/wiki/Laplace-Beltrami_operator}{https://en.wikipedia.org/wiki/Laplace-Beltrami\_operator}%
).

It is convenient to choose the constant to be $\frac{R^{2}}{4\pi}\left(
-1+\ln2\right)  $ such that%
\begin{equation}
G_{\text{Laplace}}\left(  \overrightarrow{r},\overrightarrow{s}\right)
=\frac{R^{2}}{4\pi}\ln\left(  1-\cos\Theta\right)
\end{equation}
Also note that interchange of the Laplacian with the sum gives%
\begin{equation}
\nabla^{2}G_{\text{Laplace}}\left(  \overrightarrow{r},\overrightarrow{s}%
\right)  =\sum_{\text{NB }l=1}^{\infty}\sum_{m=-l}^{l}Y_{lm}\left(
\widehat{r}\right)  Y_{lm}^{\ast}\left(  \widehat{s}\right)  =\delta
^{2}\left(  \widehat{r}-\widehat{s}\right)  -\frac{1}{4\pi}%
\end{equation}
That is to say, $\int_{S_{2}}\nabla^{2}G_{\text{Laplace}}\left(
\overrightarrow{r},\overrightarrow{s}\right)  d^{2}r=0$, no matter what the
location of $\overrightarrow{s}\in S_{2}$. \ With the interpretation that
$\nabla^{2}G_{\text{Laplace}}\propto\sigma$, the surface charge (or mass)
density, the fact that $\nabla^{2}G_{\text{Laplace}}$ integrates to zero means
the total charge (or mass) on the sphere is always zero\ (cf. Faraday's field
line interpretation). \ This \emph{fact} is quite curious and may be
surprising when first encountered. \ And while it is certainly not a big deal
for electric charge, it is rather more provocative for mass. \ It remains to
be seen if this fact requires modification of some long-established ways of
thinking about cosmology in a Newtonian framework \cite{Callan}.

In any case, we will \emph{define} the position dependent Newtonian
gravitational potential energy on the sphere, between point charges at
$\overrightarrow{r_{1}}$ and $\overrightarrow{r_{2}}$, with masses $m_{1}$ and
$m_{2}$, to be%
\begin{equation}
V\left(  \overrightarrow{r_{1}},\overrightarrow{r_{2}}\right)  =\kappa
~m_{1}m_{2}\ln\left(  1-\widehat{r_{1}}\cdot\widehat{r_{2}}\right)
\end{equation}
where $\kappa=G/R$ is the \textquotedblleft Newtonian gravitational constant
on the sphere\textquotedblright\ with units for $G$ chosen to agree with those
in 3D Euclidean space.

\subsubsection*{2-Sphere Shell theorem}

Now it should be more or less obvious that the proof of the 2D Euclidean space
shell theorem, as\ given by Newton, does \emph{not} work on the 2-sphere. \ 

For example, put a uniform ring of total mass $M$ but negligible thickness on
a circle of latitude specified by $\vartheta$. \ The Faraday points of this
system are obviously the north and south poles of the sphere. \ Next, consider
the force on an ideal point test particle, of mass $m$, located at $\left(
\theta,\phi\right)  $. \ The uniform mass density of the ring is
$\lambda=M/\left(  2\pi R\sin\vartheta\right)  $ since its radius is
$R\sin\vartheta$, thus $dM=\frac{1}{2\pi}Md\phi$ for azimuthal angular
segments of the ring. \ Therefore the net force on the test particle is \
\begin{equation}
\overrightarrow{F}\left(  \theta,\phi\right)  =F_{\theta}\widehat{\theta
}+F_{\phi}\widehat{\phi}=\frac{\kappa mM}{2\pi R}\int_{0}^{2\pi}\left(
-\widehat{\theta}\frac{\partial}{\partial\theta}\ln\left(  1-\widehat{r_{1}%
}\cdot\widehat{r_{2}}\right)  -\widehat{\phi}\frac{1}{\sin\theta}%
\frac{\partial}{\partial\phi}\ln\left(  1-\widehat{r_{1}}\cdot\widehat{r_{2}%
}\right)  \right)  d\varphi
\end{equation}
Clearly the $F_{\phi}$\ component will integrate to zero, by symmetry, but
perhaps less obviously, for $0\leq\theta\leq\pi$,
\begin{align}
f\left(  \theta,\vartheta\right)   & \equiv2\pi RF_{\theta}/\left(  \kappa
mM\right) \\
& =2\pi\left(  \frac{1}{1+\cos\theta}\operatorname{Heaviside}\left(
\vartheta-\theta\right)  -\frac{1}{1-\cos\theta}\operatorname{Heaviside}%
\left(  \theta-\vartheta\right)  \right)  \sin\theta\nonumber
\end{align}
In other words, there is in general an attractive force on the test particle,
due to this ring of total mass $M$, and that force is the same as the force
which would be produced by a \emph{point} charge of mass $M$, located at the
Faraday point in the spherical region on the other side of the ring, opposite
the test particle. \ That is to say, \
\begin{equation}
f\left(  \theta,\vartheta\right)  =f\left(  \theta,\pi\right)
\operatorname{Heaviside}\left(  \vartheta-\theta\right)  -f\left(
\theta,0\right)  \operatorname{Heaviside}\left(  \theta-\vartheta\right)
\end{equation}

To help visualize this new shell theorem, we plot $f\left(  \theta
,\vartheta\right)  $ for various ring latitudes, along with the force curves
produced by point particles at either north (i.e. $f\left(  \theta,0\right)
$) or south (i.e. $f\left(  \theta,\pi\right)  $) poles of the sphere. First,
an equatorial ring.
\vspace*{-0.5cm}
\begin{figure}[H]
  \centering
    \includegraphics[width=0.95\textwidth]{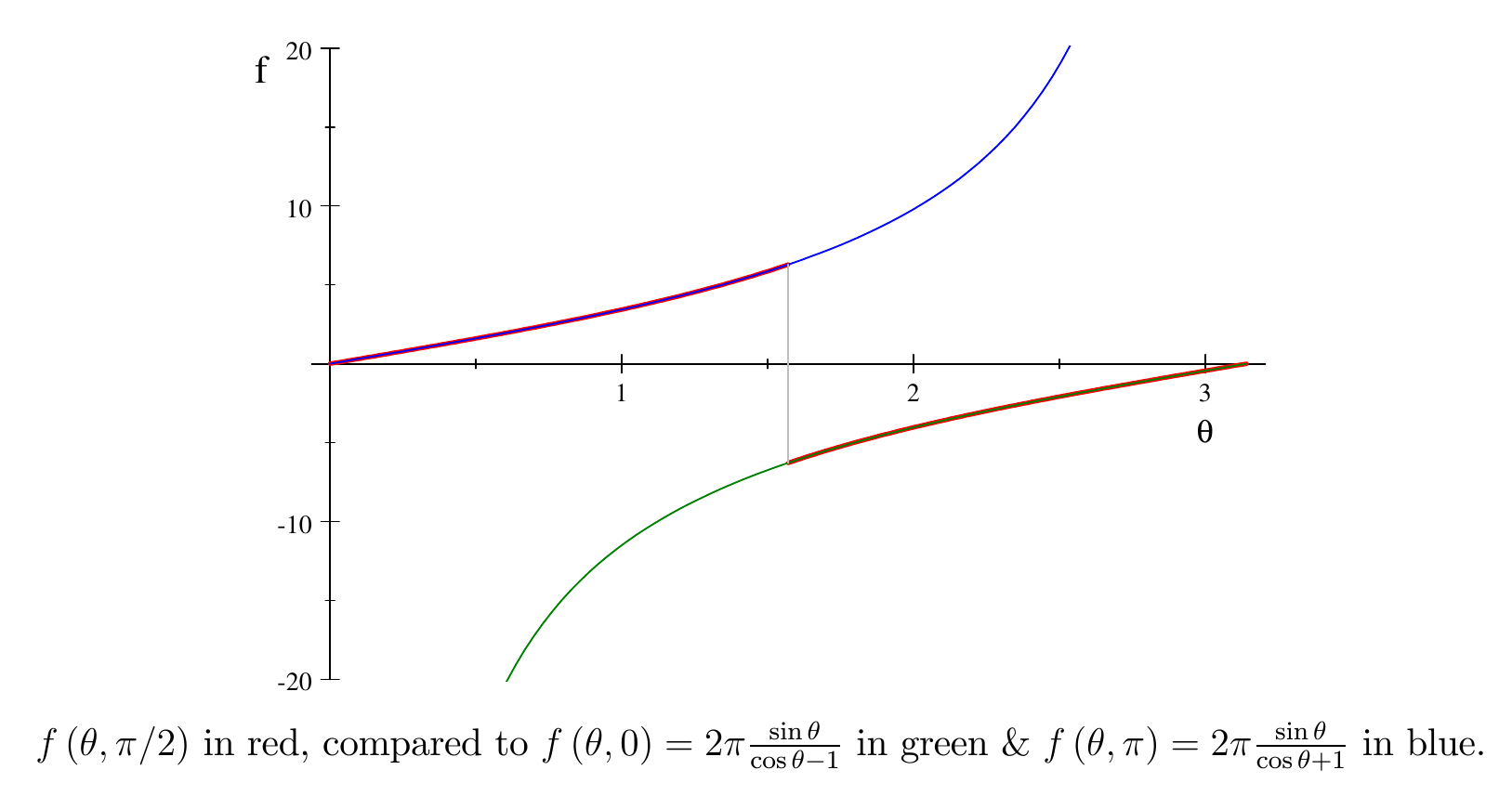}
\end{figure}
\vspace*{-0.6cm}
\par\noindent Next, a ring placed at the
\href{http://www.physics.miami.edu/~curtright/TropicOfCancer.pdf}{Tropic of
Cancer}, latitude $23^{\circ}\ 26\mrq\ 10.8"\ =\ 23.43632^{\circ} $N.
\begin{figure}[H]
  \centering
    \includegraphics[width=0.95\textwidth]{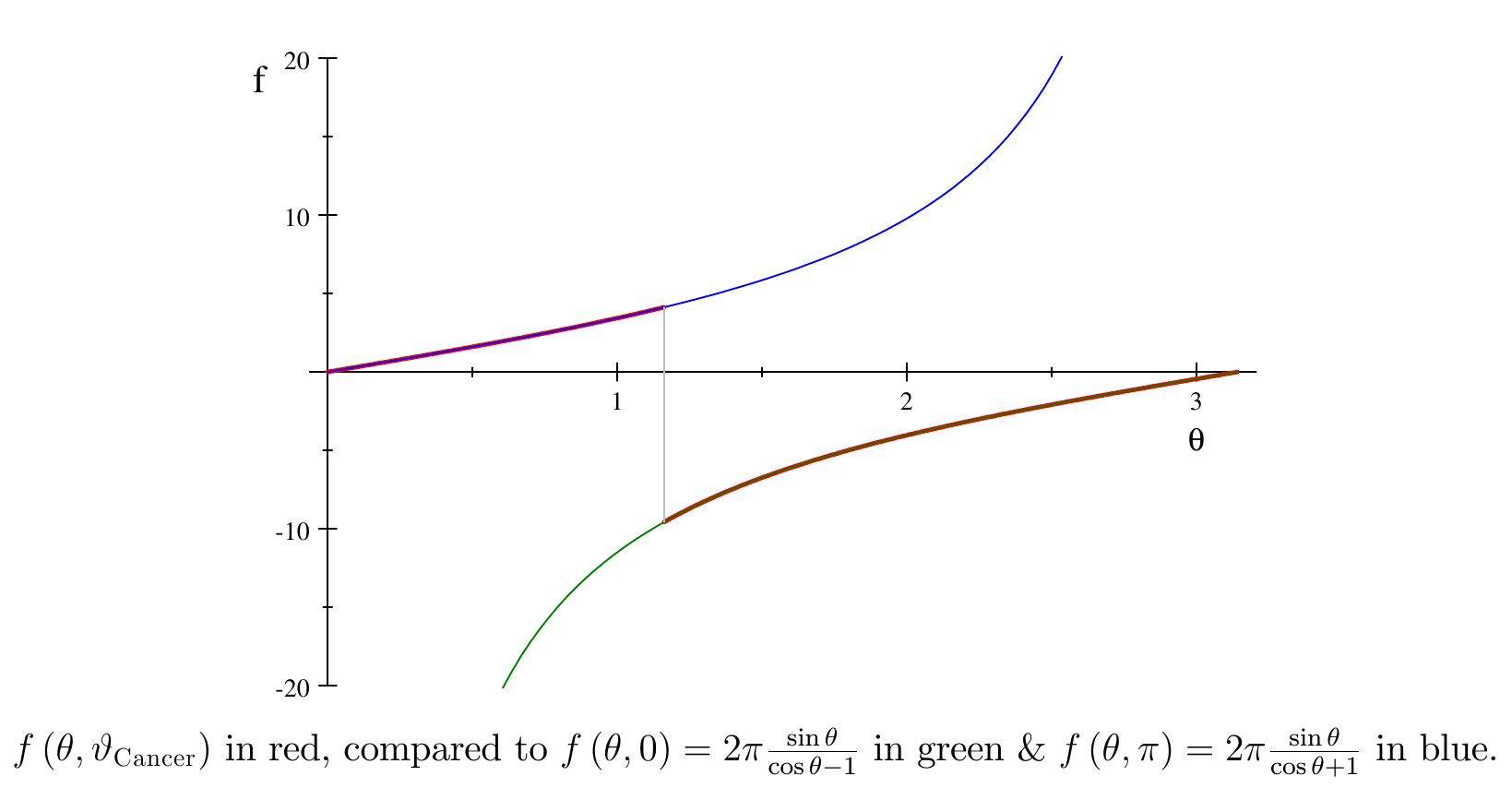}
\end{figure}
\vspace*{-0.5cm}
\noindent Note the discontinuity in $f$ at the ring's latitude\ is $4\pi
/\sin\vartheta$.\bigskip

Generalizing all this to higher N-spheres is\ now too obvious to warrant
further discussion here. \ But again, it remains to be seen if this requires
modification of some long-established ways of thinking about cosmology in a
Newtonian framework \cite{Callan}.

\subsubsection*{Miscellaneous Additional Notes for the 2-Sphere}

The perspicacious reader will have noticed that there is an obvious solution
to the inhomogeneous equation%
\begin{equation}
\frac{1}{\sin\theta}\dfrac{\partial}{\partial\theta}\left(  \left(  \sin
\theta\right)  \dfrac{\partial}{\partial\theta}h\left(  \theta\right)
\right)  =-\dfrac{1}{4\pi}%
\end{equation}
Namely,%
\begin{equation}
h\left(  \theta\right)  =\dfrac{1}{4\pi}\ln\left(  \sin\theta\right)  +const
\end{equation}
Combining with the previous result,%
\begin{equation}
\dfrac{1}{4\pi}\ln\left(  1-\cos\theta\right)  -\dfrac{1}{4\pi}\ln\left(
\sin\theta\right)  =\dfrac{1}{4\pi}\ln\left(  \tan\frac{\theta}{2}\right)
\end{equation}
Check for $\theta\neq0$: $\dfrac{\partial}{\partial\theta}\left(  \left(
\sin\theta\right)  \dfrac{\partial}{\partial\theta}\left(  \dfrac{1}{4\pi
}\left(  \ln\left(  \tan\left(  \theta/2\right)  \right)  \right)  \right)
\right)  =0$, or at least this is true so long as $\theta\neq0$. \ Also check
the Dirac delta:%
\begin{equation}
\int_{\text{cap at north pole}}\nabla^{2}\ln\left(  \tan\frac{\theta}%
{2}\right)  d^{2}\Omega=\frac{1}{R^{2}}\int_{\partial\text{cap}}\left(
\frac{\partial}{\partial\theta}\ln\left(  \tan\frac{\theta}{2}\right)
\right)  \sin\theta d\phi=\frac{2\pi}{R^{2}}%
\end{equation}
upon using $\left(  \frac{\partial}{\partial\theta}\ln\left(  \tan\frac
{\theta}{2}\right)  \right)  \sin\theta=\allowbreak1$. \ Therefore,
$\nabla^{2}\ln\left(  \tan\frac{\theta}{2}\right)  =\frac{2\pi}{R^{2}}%
~\delta^{2}\left(  \widehat{z}\right)  $, or at least that is the case near
the north pole of the sphere. \ What about the south pole?%
\begin{equation}
\int_{\text{cap at south pole}}\nabla^{2}\ln\left(  \tan\frac{\theta}%
{2}\right)  d^{2}\Omega=-\frac{1}{R^{2}}\int_{\partial\text{cap}}\left(
\frac{\partial}{\partial\theta}\ln\left(  \tan\frac{\theta}{2}\right)
\right)  \sin\theta d\phi=-\frac{2\pi}{R^{2}}%
\end{equation}
So there is another point source at the south pole. \ That is to say,%
\begin{equation}
\nabla^{2}\ln\left(  \tan\frac{\theta}{2}\right)  =\frac{2\pi}{R^{2}}~\left(
\delta^{2}\left(  \widehat{z}\right)  -\delta^{2}\left(  -\widehat{z}\right)
\right)  =\frac{4\pi}{R^{2}}~\delta^{2}\left(  \widehat{z}\right)  -\frac
{2\pi}{R^{2}}~\left(  \delta^{2}\left(  \widehat{z}\right)  +\delta^{2}\left(
-\widehat{z}\right)  \right)
\end{equation}
Subtracting the $\ln\left(  \sin\theta\right)  $ term cancelled the uniform
density on the sphere due to the $\ln\left(  1-\cos\theta\right)  $ term, but
in addition, the $\ln\left(  \sin\theta\right)  $ term contributed equal
strength Dirac deltas at both the north and the south poles, such that the
overall contribution of $\ln\left(  \sin\theta\right)  $ to the total charge
was still zero. \ Hence, the modification%
\begin{equation}
G\left(  \overrightarrow{r},\overrightarrow{s}\right)  =\frac{R^{2}}{4\pi}%
\ln\left(  \tan\left(  \frac{1}{2}\arccos\left(  \widehat{r}\cdot
\widehat{s}\right)  \right)  \right)
\end{equation}
yields a Green function whose source is only Dirac deltas, but again with
total charge zero.%
\begin{equation}
\nabla^{2}G\left(  \overrightarrow{r},\overrightarrow{s}\right)  =\frac{1}%
{2}~\delta^{2}\left(  \widehat{r}-\widehat{s}\right)  -\frac{1}{2}~\delta
^{2}\left(  \widehat{r}+\widehat{s}\right)  \propto\sigma_{\text{total}}\text{
\ \ with \ \ }\int_{S_{2}}\sigma_{\text{total}}d^{2}r=0
\end{equation}
Yet another way to look at the $\ln\left(  \sin\theta\right)  $ term is to
write%
\begin{equation}
\ln\left(  \sin\theta\right)  =\frac{1}{2}\ln\left(  \sin^{2}\theta\right)
=\frac{1}{2}\ln\left(  1-\cos^{2}\theta\right)  =\frac{1}{2}\left(  \ln\left(
1-\cos\theta\right)  +\ln\left(  1+\cos\theta\right)  \right)
\end{equation}
which reveals the two Dirac deltas more immediately.

Alternatively, one could just as well \emph{subtract} from the $\ln\left(
1-\widehat{r}\cdot\widehat{s}\right)  $ term a contribution due to a single,
equal strength Dirac delta at \emph{any} \emph{other location}. \ For example,
if the original point source is at the north pole of the sphere, put an
opposite sign point source at the south pole, to obtain%
\begin{equation}
\ln\left(  \frac{1-\cos\theta}{1+\cos\theta}\right)  =\ln\left(  \frac
{\sin^{2}\theta/2}{\cos^{2}\theta/2}\right)  =2\ln\left(  \tan\frac{\theta}%
{2}\right)
\end{equation}
Thus, if $\theta\neq0$ and $\theta\neq\pi$, once again check: $\frac{1}%
{\sin\theta}\dfrac{\partial}{\partial\theta}\left(  \left(  \sin\theta\right)
\dfrac{\partial}{\partial\theta}\ln\left(  \tan\frac{\theta}{2}\right)
\right)  =\allowbreak0$. \ Moreover,%
\begin{equation}
G\left(  \overrightarrow{r},\overrightarrow{s}\right)  =\frac{R^{2}}{2\pi}%
\ln\left(  \tan\left(  \frac{1}{2}\arccos\left(  \widehat{r}\cdot
\widehat{s}\right)  \right)  \right)  \ ,\ \ \ \nabla^{2}G\left(
\overrightarrow{r},\overrightarrow{s}\right)  =\delta^{2}\left(
\widehat{r}-\widehat{s}\right)  -\delta^{2}\left(  \widehat{r}+\widehat{s}%
\right)
\end{equation}
In fact, an arbitrary number of Dirac deltas, representing point particles
with various strengths at various locations on $S_{2}$, will eliminate the
constant charge density provided all the Dirac delta charges sum to zero.
\ Similar remarks apply for other $S_{N}$.

\subsection*{Conclusion}

We identified Newtonian and Coulombic potentials, and forces, with Green
functions on $S_{N}$, and their gradients. \ We then considered the analogue
of Newton's shell theorem for this simple system. \ We found a tasteful
generalization of the shell theorem which is simple to state. \ We believe a
similar shell theorem can be established for spheroidal and ellipsoidal if not
more general closed manifolds, as we intend to show in subsequent
work.\bigskip

\noindent\textbf{Acknowledgement:} \ This work resulted from participation in
\href{https://cgc.physics.miami.edu/BASIC/BASIC2022Summary.html}{BASIC 2022}.
\ One of us (TLC) thanks the participants of that conference for their
comments on the $N=2$ example.

\subsection*{Appendix}

The general integral expressions for the Green functions on $S_{N}$, as given
by (\ref{GreenN}) in the text, are not appropriate for the circle, $S_{1}$,
when $\theta$ is allowed to be negative with $-\pi\leq\theta\leq\pi$. \ In
that case $G$ is best obtained from first principles, with the result%
\[
G\left(  \widehat{r},\widehat{s}=\widehat{z},N=1\right)  \equiv G\left(
\theta\right)  =\frac{1}{2}\left\vert \theta\right\vert -\frac{1}{4\pi}%
\theta^{2}\ ,\ \ \ \text{for}\ \ \ -\pi\leq\theta\leq\pi
\]
The $\theta^{2}$ term eliminates a Dirac delta at $\theta=\pi$, implicit in
the second derivative of the $\left\vert \theta\right\vert $\ term, by
ensuring the slope of $G\left(  \theta\right)  $ is continuous at the south
pole of the circle, i.e. $\lim_{\theta\rightarrow\pm\pi}dG\left(
\theta\right)  /d\theta=0$. \ Thus%
\[
\frac{d^{2}}{d\theta^{2}}G\left(  \theta\right)  =\delta\left(  \theta\right)
-\frac{1}{2\pi}%
\]

Otherwise, the general integral expressions in the text are easily evaluated
for specific integer $N\geq2$. \ Here we list $G\left(  \widehat{r}%
,\widehat{s}=\widehat{z},N\right)  $ for $2\leq N\leq10$, as computed by Maple
directly from those integral expressions.%
\begin{align*}
G\left(  \widehat{r},\widehat{s}=\widehat{z},N=2\right)   & =\frac{1}{4\pi}%
\ln\left(  1-\cos\theta\right) \\
G\left(  \widehat{r},\widehat{s}=\widehat{z},N=3\right)   & =\frac{\cos\theta
}{4\pi^{2}\sin\theta}\left(  \theta-\pi\right) \\
G\left(  \widehat{r},\widehat{s}=\widehat{z},N=4\right)   & =\frac{1}{8\pi
^{2}}\left(  \ln\left(  1-\cos\theta\right)  -\frac{\cos\theta}{1-\cos\theta
}\right) \\
G\left(  \widehat{r},\widehat{s}=\widehat{z},N=5\right)   & =\frac{\cos\theta
}{8\pi^{3}\sin^{3}\theta}\left(  3\left(  \theta-\pi\right)  +2\left(
\pi-\theta\right)  \cos^{2}\theta-\sin\theta\cos\theta\right) \\
G\left(  \widehat{r},\widehat{s}=\widehat{z},N=6\right)   & =\frac{1}%
{16\pi^{3}}\left(  3\ln\left(  1-\cos\theta\right)  +\frac{\left(  4\cos
\theta-5\right)  \cos\theta}{\left(  1-\cos\theta\right)  ^{2}}\right) \\
G\left(  \widehat{r},\widehat{s}=\widehat{z},N=7\right)   & =\frac{\cos\theta
}{16\pi^{4}\sin^{5}\theta}\left(  \left(  \theta-\pi\right)  \left(
3+4\sin^{2}\theta+8\sin^{4}\theta\right)  -3\left(  \sin\theta\cos
\theta\right)  \left(  1+2\sin^{2}\theta\right)  \right) \\
G\left(  \widehat{r},\widehat{s}=\widehat{z},N=8\right)   & =\frac{1}%
{32\pi^{4}}\left(  15\ln\left(  1-\cos\theta\right)  -\frac{\cos\theta
}{\left(  1-\cos\theta\right)  ^{3}}\left(  33-54\cos\theta+23\cos^{2}%
\theta\right)  \right)
\end{align*}%
\begin{multline*}
G\left(  \widehat{r},\widehat{s}=\widehat{z},N=9\right)  =\\
\frac{\cos\theta}{32\pi^{5}\sin^{7}\theta}\left(  3\left(  \theta-\pi\right)
\left(  5+6\sin^{2}\theta+8\sin^{4}\theta+16\sin^{6}\theta\right)  -\left(
\sin\theta\cos\theta\right)  \left(  15+28\sin^{2}\theta+44\sin^{4}%
\theta\right)  \right)
\end{multline*}%
\begin{multline*}
G\left(  \widehat{r},\widehat{s}=\widehat{z},N=10\right)  =\\
\frac{1}{64\pi^{5}}\left(  105\ln\left(  1-\cos\theta\right)  \allowbreak
+\frac{\cos\theta}{\left(  1-\cos\theta\right)  ^{4}}\left(  176\cos^{3}%
\theta-599\cos^{2}\theta+696\cos\theta-279\right)  \right)
\end{multline*}

\end{document}